\documentclass[superscriptaddress,showpacs,floats, a4paper, prb, floatfix]{revtex4}
\usepackage{graphicx}
\newcommand{\cuo}{Cu$_2$O}
\newcommand{\ep}{$E_\mathrm{pump}$}

\newcommand{\xo}{$X_\mathrm{ortho}$}
\newcommand{\xod}{$X_\mathrm{ortho}^\mathrm{DI}$}
\newcommand{\xop}{$X_\mathrm{ortho}^\mathrm{PA}$}
\newcommand{\xp}{$X_\mathrm{para}$}
\newcommand{\figwidth}{7.5cm}
\newcommand{\sfigwidth}{6cm}

\begin{document}
\title{Decay and coherence of two-photon excited yellow ortho-excitons in \cuo.}
\author{K. Karpinska}
\altaffiliation{Present address: Institute of Physics, Polish
Acade\-my of Sciences, 02 668 Warsaw, Poland. e-mail: karpik@ifpan.edu.pl}
\author{M. Mostovoy}
\author{M.A. van der Vegte}
\affiliation{Materials Science Center, University of Groningen,
Nijenborgh 4, 9747 AG Groningen, The Netherlands}
\author{ A. Revcolevschi}
\affiliation{Laboratoire de Physico-Chimie de l'Etat Solide,
Universit\'e Paris XI, France}
\author{P.H.M. van Loosdrecht}
\email{p.h.m.van.loosdrecht@rug.nl}
\affiliation{Materials Science Center, University of Groningen,
Nijenborgh 4, 9747 AG Groningen, The Netherlands}
\date{\today}

\begin{abstract}
Photoluminescence excitation spectroscopy has revealed a novel,
highly efficient two-photon excitation method to produce a cold,
uniformly distributed high density excitonic gas in bulk cuprous
oxide. A study of the time evolution of the density, temperature
and chemical potential of the exciton gas shows that the so called
quantum saturation effect that prevents Bose-Einstein condensation
of the ortho-exciton gas originates from an unfavorable ratio
between the cooling and recombination rates. Oscillations observed
in the temporal decay of the ortho-excitonic luminescence
intensity are discussed in terms of polaritonic beating. We
present the semiclassical description of polaritonic oscillations
in linear and non-linear optical processes.
\end{abstract}

\maketitle

\section{Introduction}
\label{sec:introduction} The quest for Bose-Einstein
condensation (BEC) of excitons has been one of the main
driving forces behind the numerous studies on high density
excitonic gases during the last few
decades.\cite{mos00,gri95} A variety of systems have been
tried, including bulk crystals\cite{gri95,sno00} and
quantum well systems.\cite{lai04,sno03} In spite of the
intense effort there still is no clear evidence that
condensation was actually reached. The formation of a
condensate is expected to occur if the Bose gas reaches a
quasi-equilibrium state characterized by a sufficiently
high density $n$ and, simultaneously, a sufficiently low
temperature $T$. For a three dimensional non-interacting
Bose gas in equilibrium the transition temperature $T_c$
is proportional to $n^{2/3}$. Strong spatial confinement
and decreased dimensionality lead to a substantial
decrease of the condensation temperature. This is why bulk
crystals remain very attractive in the search for excitonic
BEC. The long life times ($>1$~ns) and high binding energy
($150$~meV) of the excitons in \cuo\ make cuprous oxide an
excellent candidate in this class. In addition, the small
Bohr radius ($a_B\approx7$~\AA) of the ground state
exciton assures that even at densities as high as
$n\approx10^{20}$~cm$^{-3}$ the exciton-exciton
interactions remain weak.\cite{sno91}
Apart from the intriguing possibility of BEC, there is a number of
other interesting  phenomena and properties of the exciton gas in
\cuo\ which attracted attention, including localization of
excitons in a stress trap,\cite{nak02,nak04} spectroscopy on
intra-excitonic transitions,\cite{kub05} and the observation of
polariton beating in transmission experiments resulting from
interference of excited states belonging to different polariton
dispersion branches.\cite{Froelich,fro91}
In this paper we revisit the high density ortho exciton gas in
\cuo. The first part of the paper is concerned with spectroscopic
photoluminescence excitation  experiments revealing a
novel and efficient non-linear exciton excitation scheme, which
has several advantages over the schemes used so far. In addition,
we present a study of some of the statistical properties of the
thus created exciton gas, and comment on the origin of the so
called ``quantum saturation effect''\cite{sno87} that prevents BEC
in the ortho-exciton gas in cuprous oxide. The second part of the
paper concentrates on the polaritonic beating effect observed in
our experiment for which a novel analytical model will be
introduced. The new model explains both the beating reported
earlier\cite{Froelich,fro91} for the transmission geometry as well
as our results obtained for the back-reflection configuration.

\begin{figure}[ht]
\centering\includegraphics[width=\figwidth]{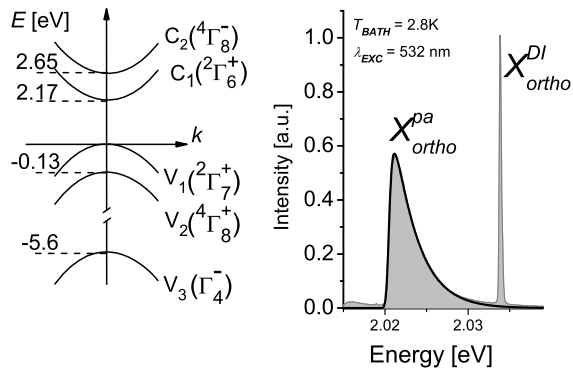}
\caption{Left panel: Schematic energy band diagram at the $\Gamma$
 point in \cuo. Right panel:
 Ortho-exciton emission spectrum at $T=2.8$~K using 532 nm
 continuous wave excitation. The thick line represents a fit to
 Eq.(\ref{response}) using a Maxwell-Boltzmann distribution with
$T_\mathrm{be}=24.7$~K. \label{band}}
\end{figure}

Cuprous oxide, \cuo, has four different excitonic series. These
are attributed to various combinations of valence ($V$) band holes
and conduction ($C$) band electrons (see Fig.~\ref{band}):
$V_1-C_1$ (yellow),$V_2-C_1$ (green), $V_1-C_2$ (blue) and
$V_2-C_2$ (indigo). The two uppermost valence bands originate from
3$d$ copper orbitals (split by spin-orbit interaction) while the
lowest conduction band is mainly formed from 4$s$ copper
orbitals.\cite{rui97} Hence all three bands ($V_1$, $V_2$ and
$C_1$) have the same even parity, and an optically forbidden
direct gap (E$_g=2.17$~eV at 10~K) exists at a center of a
Brillouin zone. The lowest excitonic state of the system is split
by the electron-hole exchange interaction into a 1$s$-ortho state
($J=1$) at 2.033~eV (\xo) and a 1$s$-para state ($J=0$) at 2.021
eV (\xp). As a consequence of the even parity of the bands
involved, the efficiency of ``direct" optical pumping of the
yellow excitons is very low. Radiative decay of 1$s$
ortho-excitons is weakly allowed through a quadrupolar coupling.
For the 1$s$-para-exciton radiative transitions are forbidden to
high order.

The yellow ortho-exciton emission spectrum consists of two main
features: a direct $k\approx0$ transition (\xod) at 2.034~eV, and a
phonon-assisted (\xop) emission band at 2.02~eV (see
Fig.\ref{band}). The PA-band results from decay by simultaneous
emission of a photon and a LO-phonon of energy 13.6~meV.
Since both the matrix element and the phonon dispersion are nearly
$k$-independent\cite{oha00} the shape of
the PA band directly reflects the kinetic energy distribution of
the ortho-exciton gas. The observed emission band may therefore
be described by
\begin{equation}
I(E)=A\int^{\infty}_{-\infty}D(x)f(x)e^{(\frac{x-E}{\Gamma})^2} dx \ ,
\label{response}
\end{equation}
where $D(E)\propto E^{1/2}$ is the density of excitonic states
(assuming a quadratic dispersion), and $f(E)$ is either the
Maxwell-Boltzmann ($f_\mathrm{mb}$, classical gas) or
Bose-Einstein ($f_\mathrm{be}$, quantum gas) distribution
function. The product $D(x)f(x)$ is convoluted with a Gaussian
experimental resolution function of width $\Gamma$. Fitting this
equation to the experimentally observed spectrum yields an
estimate for the effective excitonic temperature and chemical
potential ($\mu$). For an ideal Bose gas  above the
condensation temperature, the density of bosons is given by
\begin{equation}\label{density}
n=g(m^*/2\pi \hbar^2)^{3/2} \int f_{\mathrm{be}}(E,\mu,T)D(E)dE\ ,
\end{equation}
with $g$ the degeneracy, and $m^*$ the excitonic mass.

The legitimacy of this analysis has been questioned in
Ref.~\onlinecite{oha00}, where it was argued that Auger processes
might lead to a serious overestimation of the exciton density.
Subsequent experimental studies,\cite{den02} however, have shown
that this is only relevant near the surface. The experiments
described here use excitation in the near infrared, where the
optical penetration depth is larger than 500~$\mu m$.\cite{teh83}
Any contribution due to surface enhanced Auger processes is
therefore expected to be negligible. For the same reason (large
penetration depth) the thus created exciton gas in nearly
uniformly distributed in the sample. Therefore effects
related to a strong spatial inhomogeneity of the gas can also be
neglected in our case.
\section{Photoluminescence excitation spectroscopy}
\label{sec:luminescence} The samples used in this study were [100]
platelets cut from a single crystal grown by a floating zone
technique.\cite{sch74} The upper limit for the full width at half
maximum of the ortho-exciton direct emission at T=10K is measured
to be 0.09nm (0.3meV) which is an experimental resolution limited
result. The excitonic emission is observed up to the room
temperature. In the emission spectrum obtained for overbandgap
excitation ($E>2.17$~eV) a broad band is observed peaked
at 1.7~eV attributed to
ionized oxygen vacancy $V_O^{2+}$.\cite{Ohyama} The relative
intensity of the (\xo) and $V_O^{2+}$ bands is comparable to those
reported for high quality samples and shows that the samples used
in this study have a relatively low density of the oxygen
vacancies. Emission peaks attributed to excitons bound to
impurities that have been reported around 2eV are absent in our
spectra.\cite{Ohyama} Moreover we observe an exceptionally long
lifetime of paraexcitons in our samples, $\tau \approx
14$~ms,\cite{Karp} independently demonstrating the high quality of
the sample since the decay of paraexcitons is exclusively defect
limited.

For the photoluminescence excitation experiments
samples were polished and mounted on the cold finger of a
continuous flow cryostat and kept at $T=4$~K. The time integrated
experiments were performed using an
optical parametric amplifier pumped by the third harmonic of a 10
Hz Nd:YAG laser. Excitation pulses with photon energies in the $1-2.5$~eV range and
temporal width of 8~ns were weakly focused on the sample (spot
size 500~$\mu$m). The yellow exciton emission was detected in
a back-reflection geometry using a 0.5~m double-grating
monochromator in combination with a photomultiplier.

\begin{figure}[ht]
\centering\includegraphics[width=\figwidth]{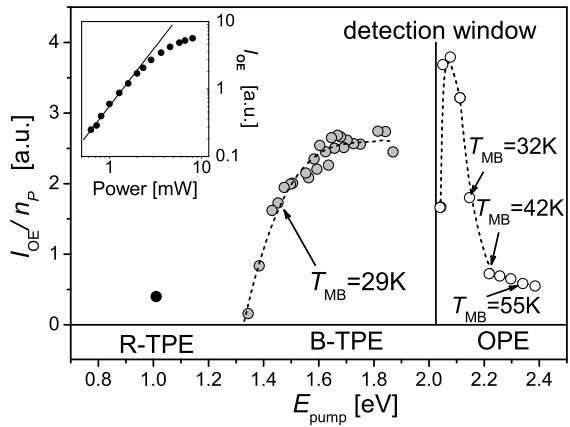}
\caption{Integrated intensity of the ortho-exciton emission per
incoming photon ($I_{OE}/n_p$) versus the pump laser energy (\ep).
Bath temperature $T=7$~K, $P=3\ \mathrm{mW}=\mathrm{constant}$.
Full circles - Blue-exciton Two Photon Excitation (B-TPE), open
circles - One Photon Excitation (OPE) and black dot - 1$s$
Resonant Two-Photon Excitation (R-TPE). Dashed lines are guides to
the eye. Inset: $I_{OE}$ as a function of the pumping laser power
for \ep$=1.4$~eV. Solid line shows a fit for $I\sim P^{\alpha}$
with $\alpha=1.8$.} \label{excitation}
\end{figure}

Figure~\ref{excitation} shows the
results of a photoluminescence excitation study of yellow exciton
creation in \cuo. During the experiments the pump fluency was kept
constant at 30 W/cm$^2$.
The integrated intensity $I_\mathrm{OE}$ of the ortho-exciton
emission (both phonon-assisted and direct) is
normalized by the number of incoming photons per second
$n_\mathrm{p}$, and is therefore a measure of the efficiency of
the pumping.

Optical pumping is the most commonly used method to
produce excitons. The efficiency of creating the excitonic
gas will naturally depend on the particular process
involved, and hence on the pump photon energy \ep. The
observed excitation spectrum may be divided into several
regions representing different excitation processes. A
first rough division can be made into a one-photon (OPE)
and a two-photon (TPE) excitation region, with the divider
at the 1s exciton energy. In the high energy OPE region
(\ep$>E_\mathrm{gap}\approx2.17$~eV) the dominant process
is direct excitation of electron-hole ($C_1-V_1$) pairs,
which may subsequently form an exciton. As mentioned
above, however, this process is not very effective due to
the even parity of both bands involved. For \ep$<E_{gap}$
the process becomes more efficient since single photon
transitions to excited yellow exciton states as well as
resonant Raman transitions are allowed. The OPE efficiency
reaches its maximum when \ep~ approaches \xod+LO at
2.05~eV~\cite{gor00} and decreases gradually upon
increasing excitation energy. It agrees with the fact that
the absorption coefficient at the excitonic states of
higher quantum number $n$ decreases as $\approx 1/n^{3}$.
Decrease of the number of free excitons created by pumping
photon may also be related to the fact that overbandgap
excitation enables retrapping of either electrons or
hot-excitons by defect sites.\cite{Ito} Moreover a surface
photo-voltage effect may take place, which leads to a
space-charge accumulation at the surface.

Resonant two photon excitation (R-TPE, black data point at
1.015 eV) to the ortho-exciton state is allowed by
symmetry and could thus, in principle, be efficient.
Unfortunately, the matrix element for this process is
fairly small since it involves an intermediate state which
lies as deep as 5.6~eV below the upper valence band
($V_1$),\cite{rus73} in agreement with the low efficiency
observed.

The so far unexplored part of the excitation spectrum is
the 1.3-1.9~eV region. Surprisingly efficient excitation
of the 2.036 eV yellow ortho-excitons is observed for \ep\
significantly below \xo\ but still well above \xo/2. The
precise nature of the process responsible will be
discussed later on. At this point, it is sufficient to
note that it is a two photon process with $V_1-C_2$
electron-hole pairs as intermediate state. This is
evidenced by the observation of a threshold for the
process at $\sim1.3$~eV which coincides with half the
$V_1-C_2$ energy gap and by the near quadratic dependence
of the excitonic emission on the fluency (see inset
Fig.~\ref{excitation}) measured for \ep=1.4eV.

The origin of the saturation of the efficiency for high
fluencies ($P>>40 W/cm^2$) remains to be investigated. It
has also been observed for other excitation methods, and
may originate from two-exciton spin-flip
processes.\cite{kav02}

For the quest for Bose-Einstein condensation in excitonic matter
it is vital to have an efficient excitation process ({\em i.e.} a
high density). The limited lifetime of the exciton gas, however,
also imposes the requirement of a low initial exciton gas
temperature. At first sight, this seems problematic for the new
B-TPE process. The process has an excess energy of at least 0.5
eV, which might be dissipated in either the exciton gas or in the
lattice. A first estimate of the temperature of the gas can be
made by fitting Eq.(\ref{response}) to the PA band (see
Fig.~\ref{band}), using the Maxwell-Boltzmann distribution. The
thus obtained gas temperatures are denoted in
Fig.~\ref{excitation} for several OPE and B-TPE excitation
energies. In all cases the exciton temperature is higher than the
bath temperature (4~K). In the OPE region, the smaller the energy
mismatch between \ep\ and \xo, the lower $T_\mathrm{mb}$. It
decreases from 55~K for \ep$=2.3$~eV to 32~K for 2.15~eV.
Surprisingly, an even lower gas temperature is observed when
$1.4<E_\mathrm{pump}<1.9$~eV in spite of the fact that the energy
mismatch (0.1-0.7~eV) is much larger than in the OPE cases. The
exciton temperature in this range is found to be 29 K, nearly
independent on the photon energy. Apparently, the excess energy
leaves the system. One possibility would be that the $V_1-C_2$
electron-hole state decays to the yellow exciton state through a
dipole allowed transition either direct, or via  an intermediate
$V1-C1$ electron-hole state. Another possibility is that the
$V_1-C_2$ state acts as an intermediate state for a hyper-Raman
type process. In both cases the excess energy is emitted as a
photon, which, in view of the low absorption coefficient at low
energy, would indeed leave the sample. For determination of the
precise nature of the process involved, one would need to perform
emission experiments in the energy range 0.1-0.7 eV.
\section{time resolved two photon spectroscopy}
\label{sec:timeresolved}
A more detailed insight into the
properties of the B-TPE created exciton gas may be
obtained from time-resolved luminescence experiments. Time
resolved measurements were performed using a traveling
wave optical parametric amplifier pumped by 1~kHz,
1.55~eV, 120~fs pulses produced by an amplified
Ti-Sapphire laser system. The 150~fs excitation pulses
(1.3-2.5~eV) were focused on the sample with a spot size
of about 100$\mu$m. The exciton emission was detected by a
Hamamatsu streak camera system operating in photon
counting mode (temporal resolution 10 ps). The
experimental spectral resolution function is measured to
be of a Gaussian type with $\Gamma$=0.37nm.

\begin{figure}[ht]
\begin{center}
\includegraphics[width=\figwidth]{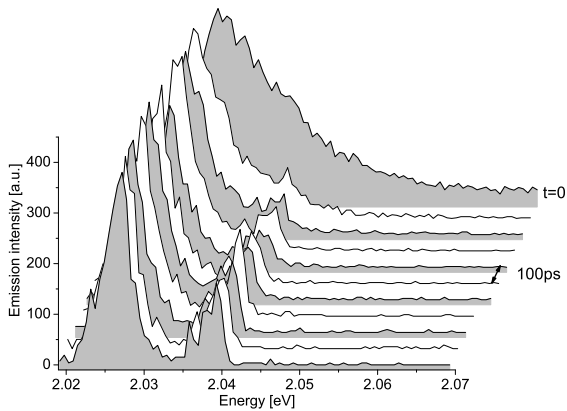}
\end{center}
\vspace*{-0.5cm} \caption{Time evolution of the ortho-exciton
emission spectrum after excitation with a 150~fs, \ep$=1.4$~eV pulse.} \label{waterfall}
\end{figure}

Figure~\ref{waterfall} shows the time-resolved ortho-exciton
luminescence spectra after excitation with a 150 fs pulse
(\ep$=1.4$~eV) at $T=7$~K. Initially, a broad spectrum is observed
dominated by the phonon assisted transitions reflecting the
relatively high excitonic temperature. Despite the rather indirect
way of excitation, no initial growth of the luminescence has been
observed, and the exciton creation appears to be instantaneous
upon excitation within the experimental resolution (10 ps). As
time progresses both the intensity and the shape of the emission
change, where the former reflects the cooling of the exciton gas,
and the latter the temporal decay of the exciton population.
Direct and phonon-assisted peaks become well resolved for delay
times longer than 0.2~ns, and show a similar long time decay
dynamics.
\begin{figure}[ht]
\centering\includegraphics[width=\sfigwidth]{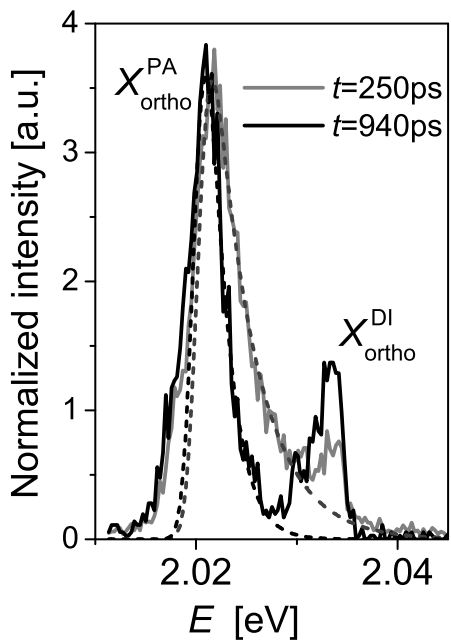}
\caption{Left panel: Schematic energy band diagram at the $\Gamma$
 point in \cuo. Right panel:
 Time resolved ortho-exciton emission spectra at $t=250$~ps and $t=940$~ps.
 Dashed lines are a fit to Eq.(\ref{response}) using Bose-Einstein statistics
 with $T_{\mathrm{be}}=64\ \mathrm {K};\
 \mu=-0.7\ \mathrm{meV}$, and $T_\mathrm{be}=23\ \mathrm {K};\
 \mu=-0.24\ \mathrm{meV}$, respectively.} \label{trspec}
\end{figure}

Two representative time-resolved \xo\ luminescence spectra are
shown in figure~\ref{trspec}, recorded 250 and 950 ps after
excitation with a 150 fs pulse with \ep$=1.4$~eV, together with
fits to Eq.(\ref{response}), using a Bose-Einstein distribution
function. The values for the gas temperatures at these times are
found to be $T=34$~K, and $T=17$~K, respectively. As one can see,
the fits reproduce the high energy tail very well. The deviations
observed at the onset of the emission are due to replica-like
structures which presumably appear due to a small strain induced
by the mounting of the sample on the cold finger of the
cryostat.\cite{sun02} In the final fitting of the data this was
taken into account by adding a scaled down replica of the
spectrum, shifted by about 3 meV to Eq.(\ref{response}).

\begin{figure}[ht]
\centering\includegraphics[width=\figwidth]{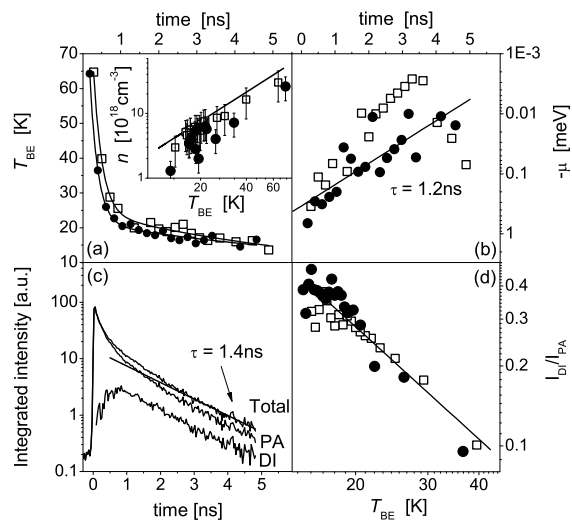} \caption{ Upper
panels: Time dependence of (a) the exciton temperature
$T_\mathrm{be}$ and (b) chemical potential. Solid lines are fits
to a two- and one-exponential decay, respectively. The inset in
(a) shows the $n,T$ phase diagram (solid line is the BEC
boundary), where $n$ has been calculated using Eq.(\ref{density}).\\
Lower panels: (c) Temporal decay of the integrated intensities of
the total, direct (DI), and phonon assisted (PA) luminescence.
Solid line: fit to a single exponential decay. (d) Ratio between
direct and phonon assisted luminescence as function of the gas
temperature, solid lines are a guide to the eye. The data sets are
for \ep$=1.4$~eV (circles), and \ep$=2.17$~eV (squares).
\label{cooling}}
\end{figure}

The parameters resulting from fitting Eq.(\ref{response}) to time
resolved spectra of the phonon assisted luminescence are shown in
Fig.~\ref{cooling} (upper panels). The two parameter sets shown
correspond to experiments with excitation with 1.4 eV (filled
symbols) and 2.17 eV (open circles), {\it i.e.} to B-TPE and OPE
excitation, respectively. The cooling of the gas, shown in
Fig.~\ref{cooling}a), shows a bi-exponential behavior. Initial
fast cooling ($\tau\sim0.2$~ns) occurs via optical phonon
emission. Once the kinetic energy of the excitons becomes too low
for this process, the relaxation occurs via acoustical phonon
emission with a relatively long decay time ($\tau\sim6$~ns).
Though the cooling rate hardly depends on the excitation method,
it is clear from the figure that B-TPE excitation yields lower gas
temperatures due to the lower initial temperature. The chemical
potential, shown in Fig.~\ref{cooling}b), decays faster than the
temperature, with a time constant $t\sim1.2$~ns. This decay is
essentially due to loss of particles. Consistently, the integrated
intensity (Fig.~\ref{cooling}c) of the luminescence shows a
similar decay time ($t\sim1.4$~ns). The conversion of
ortho-excitons into para-excitons, the main cause of the loss of
particles, strongly depends on temperature, hence the
non-exponential behavior at early times.\cite{den02} The
observation that the particle decay is faster than the cooling
rate, makes it implausible that the gas would undergo a BEC
transition. This is demonstrated in the inset of
Fig.~\ref{cooling}a), which shows the $(n,T)$ phase diagram. The
state of the gas follows the BEC phase boundary, and no crossing
is observed, similar to earlier observations.\cite{sno87}
Apparently, the gas adjusts its quantum properties to the density
and temperature and is in quasi-equilibrium.

As the gas cools down the relative occupation of the $k\approx 0$
state is expected to increase. Indeed, the direct emission
initially increases (see Fig.~\ref{cooling}c), and the ratio of
direct to phonon assisted emission increases with time (decreases
with temperature, Fig.~\ref{cooling}d). For B-TPE excitation this
ratio is higher than for OPE excitation for given $T_{BE}$. One
can therefore conclude that the B-TPE process leads to a
preferential occupation of the $k\approx 0$ state, which in itself
is favorable for BEC condensation.

Different explanations have been proposed for the  so-called
``quantum saturation'' effect. In Ref.~\onlinecite{Lee} it has
been suggested that the quantum saturation phenomenon originates
from the fact that the ortho-gas is not in an equilibrium state so
that the number of ortho-excitons is not well defined. As a
consequence $\mu=0$ by definition and no phase transition is
possible. The present results do not support this conclusion.  The
kinetic energy distribution of the ortho-gas can be described with
a very good accuracy by the BE distribution, and moreover decay of
the chemical potential coincides with the particle decay obtained
independently from the emission intensity decay. Therefore, one
may conclude that the ortho-gas is in fact in quasi-equilibrium,
and the particle density is a well defined number. Apparently the
exciton gas adjusts its  properties even though the phase
boundary is not crossed. Kavoulakis\cite{kav02} has argued  that
the above mentioned fitting based analysis to extract the exciton
temperature and density is misleading since the ortho-gas is not
in a quantum state but in a normal state and under this assumption
the gas slides along adiabats given by $n/n_Q\sim0.01$, where
$n_Q$ is the quantum concentration. However, for this to be true,
it would require an overestimation of the exciton density by two
orders of magnitude, and it does not explain how a classical gas
can show a non-classical kinetic-energy distribution. A few
possible explanations for an apparent non-classical distribution
have, however, been given in literature: dominant Auger
recombination and/or strong spatial inhomogeneity. As has been
discussed in section I, both effects are expected to have only a
marginal influence on the decay kinetics for B-TPE excitation.
Since for OPE pumping similar results are found, the significance
of these two effects is questionable in that case as well.
Finally, Ell {\it et al.}\cite{ell} presented an analysis of the
ortho-exciton relaxation which included polaritonic effects. The
main conclusion of that analysis is that the so called polaritonic
``bottleneck effect'' prevents formation of an ortho-excitonic BEC
at k=0. However, as pointed out in Ref.~\onlinecite{haug}, the
authors assumed in their model an infinite volume and it is not
clear to what extend their conclusions would change due to finite
volume corrections. From the above analysis we can conclude that
only the third model does not contradict with the present results
but the full model is still to be established.

Aware of the limitations mentioned in the introduction, we applied
the same analysis to \ep=2.17~eV data, where the penetration depth
is only $30\ \mu$m, and the results lead to similar conclusions as
in the B-TPE case. The fact that the early time density reached
for \ep=2.17~eV excitation is significantly higher than for
\ep=1.4~eV excitation agrees well with the $\sim$10 times smaller
penetration depth, and the 5 times lower pumping efficiency. This
indirectly supports the use of Eq.(\ref{density}), even in the
case of a small penetration depth.

\section{polariton oscillations}
\label{sec:oscilations} Analysis of the \xo\  emission presented
in the previous sections shows that absorption of the pump photons
results in the formation of ortho-exciton through a fast and
efficient process. The intensities of both \xop\ and \xod\
emission decay mainly as a result of down-conversion of the
ortho-excitons into para-excitons. More detailed analysis of the
\xod\ however reveals also weak oscillations in the measured \xod\
emission intensity ($I_{od}$), reminiscent of those previously
observed  by Fr\"{o}hlich  et al.\cite{Froelich,fro91}
\begin{figure}[ht]
\centering\includegraphics[width=\figwidth]{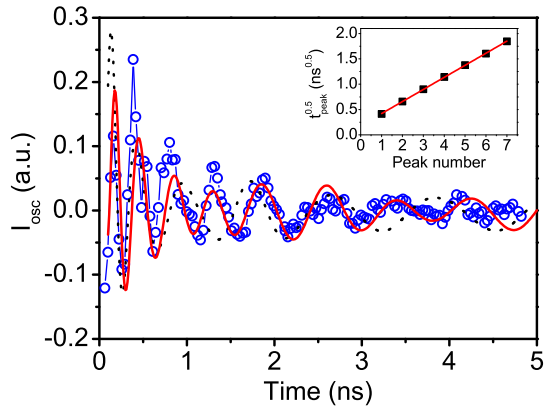} \caption{
Direct emission versus time from figure 5(c) after subtraction of
a single exponential decay curve (symbols). Dashed and
solid lines: theoretical simulation as discussed in
section\ref{sec:nonlinear}. The inset shows the $\sqrt{t}$
dependence of the peak positions. \label{oscillations}}
\end{figure}
This is shown in Fig.~\ref{oscillations}, which displays the time
dependence of the intensity after subtraction of a smooth
exponential decay curve. The oscillations have a typical
``period'' of about 0.5 ns, which increases with time as
$\sqrt{t}$ (see also inset fig.\ref{oscillations}). They are
thought to originate from interference between states of the lower
and upper polaritonic branches.\cite{Froelich,fro91} Such
oscillations have so far only been observed for resonant 
one-photon excitation in a transmission geometry, and it is not
immediately clear that the polaritonic interference would be
relevant to the present two-photon experiment and a B-TPE
excitation scheme. In the following section we therefore make an
excursion to the semiclassical theory of the polariton
oscillations of transmitted light, which allows us to obtain an
analytical expression for these oscillations. Most importantly,
the results obtained in Sec.~\ref{sec:transmission} can be
generalized to describe more complicated processes, such as the
present time resolved two photon experiment. This is done in
Sec.~\ref{sec:nonlinear}, where we show that the aperiodic
oscillations presented in Fig.~\ref{oscillations} indeed result
from the polariton beating.

\section{Polariton oscillations in transmission of light
and geometrical optics} \label{sec:transmission}

In this section we give a semiclassical description of
the polariton oscillations, observed in the transmission
experiment on Cu$_2$O.\cite{Froelich,fro91} In principle,
the transmission intensity can be found by a
straightforward numerical calculation, as was done in
Ref.[\onlinecite{Froelich}]. Here we obtain an accurate
analytical expression for this intensity.

\begin{figure}[htbp]
\centering\includegraphics[width=\sfigwidth]{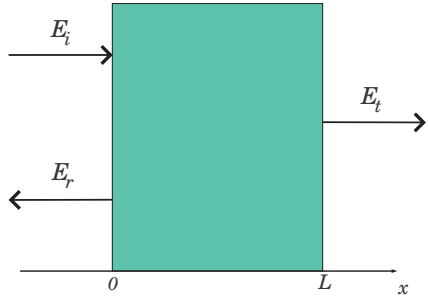}
\caption{\label{fig:transmission} The transmission of
light propagating in the $x$ direction through the slab.}
\end{figure}

\begin{figure}[htbp]
\centering\includegraphics[width=\figwidth]{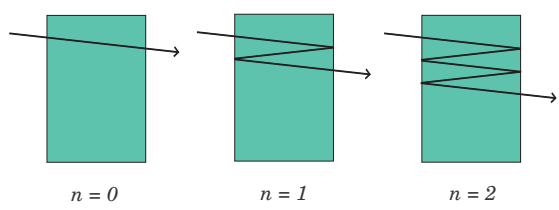}
\caption{\label{fig:multiple} Examples of processes with different
numbers of internal reflections.}
\end{figure}

Consider the propagation of linearly polarized light with
frequency $\omega$ through a slab of a material with
dielectric function $\epsilon $ (see
Fig.~\ref{fig:transmission}). For a monochromatic incident
electric field $E_{i}=E_{0}e^{-i\omega t+ikx}$, where
$k=\frac{\omega }{c}$, the amplitude of the transmitted
field is
\begin{equation}
E_{t}=E_{0}e^{-i\omega t+iqL}\frac{\left( 1-R^{2}\right)}
{1-R^{2}e^{2iqL}}, \label{Eout1}
\end{equation}
where $L$ is the slab thickness, $q=\frac{\omega }{c}\sqrt{\epsilon
}$, and $R=\frac{\sqrt{\epsilon }-1}{\sqrt{\epsilon }+1}$ is the
reflection amplitude. Equation (\ref{Eout1}) allows for a simple
interpretation: the electric field is multiplied by the factor $1-R$
when the light enters the material, and by $1+R$ when it gets out,
$-i\omega t+iqL$ is the phase acquired by the light traveling through
the sample during the time $t$, and the denominator
\[
\frac{1}{1-R^{2}e^{2iqL}}=1+R^{2}e^{2iqL}+R^{4}e^{4iqL}+\ldots
,
\]
is the sum over the number of internal reflections from the
boundaries the slab (see Fig.~\ref{fig:multiple}): a factor $R$
for each internal reflection and $e^{iqL} $ for the propagation
from one side of the slab to the other side.

Consider now the incident electric field having the form of the
Gaussian wave packet with the average frequency $\bar{\omega}$ and
width $\Delta $:
\[
E_{in}=\frac{E_{0}}{\sqrt{2\pi }\Delta} \int d\omega e^{-i\omega
t+ikx-\frac{\left( \omega -\bar{\omega}\right) ^{2}}{2\Delta ^{2}}}.
\]
Then the transmitted field at $x=L$ is given by
\begin{equation}
E_{t} =\frac{E_{0}}{\sqrt{2\pi }\Delta }\sum_{n=0}\int
d\omega \left( 1-R^{2}\right) R^{2n}e^{iS_{n}-\frac{\left( \omega -\bar{%
\omega}\right) ^{2}}{2\Delta ^{2}}},  \label{Eout2}
\end{equation}
where the action $S_{n}=-\omega t+q\left( 2n+1\right) L$ is the
phase of the field for a path with $2n$ internal reflections.
Since $L$ is usually much larger than the wavelength, the integral
in Eq.(\ref{Eout2}) can be calculated in the stationary phase
approximation:
\begin{equation}
E_{t} \approx E_{0}\sum_{n=0}\frac{\left(
1-R^{2}\right) R^{2n}}{\sqrt{1-i\left( 2n+1\right) L\Delta ^{2}\frac{d^{2}q}{%
d\omega ^{2}}}}e^{iS_{n}-\frac{\left( \omega _{n}-\bar{\omega}\right) ^{2}}{%
2\Delta ^{2}}},  \label{Eout3}
\end{equation}
where the values of $S_{n}$, $\frac{d^{2}q}{d\omega ^{2}}$, and $R$
are taken at the `stationary' frequency $\omega _{n}$, such that
\begin{equation}
\frac{dS_{n}}{d\omega }\left( \omega _{n}\right) +i\frac{\left( \omega _{n}-%
\bar{\omega}\right) }{\Delta ^{2}}=-t+\frac{dq}{d\omega
}\left( 2n+1\right) L+i\frac{\left( \omega
_{n}-\bar{\omega}\right) }{\Delta ^{2}}=0. \label{omegan1}
\end{equation}
For a broad wave packet, the term $\propto \frac{1}{\Delta
^{2}}$ can be neglected and we obtain
\begin{equation}
\left. \frac{d\omega }{dq}\right| _{\omega =\omega
_{n}}=\frac{\left( 2n+1\right) L}{t},  \label{omegan2}
\end{equation}
which is the geometrical optics result: the group velocity
of light in the slab equals the path $\left( 2n+1\right)
L$ divided by the time of motion $t$. Furthermore, for a
broad wave packet
\begin{equation}
E_{t} \approx E_{0}\sum_{n=0}\frac{\Delta _{n}}{%
\Delta }\left( 1-R^{2}\right) R^{2n}e^{i\left(
S_{n}+\frac{\pi }{4}\right) }, \label{Eout4}
\end{equation}
where $\Delta _{n}=\left[ \left( 2n+1\right) L\frac{d^{2}q}{d\omega ^{2}}%
\right] ^{-1/2}\ll \Delta $ is the width of the frequency interval around $%
\omega _{n}$ that gives the dominant contribution to the integral
in Eq.( \ref{Eout2}).

If there are no excitations with $\omega \sim
\bar{\omega}$ present in the sample, $\epsilon $ is a slow
function of frequency and can be replaced by a constant
$\epsilon _{0}=\epsilon _{0}^{\prime}+i\epsilon
_{0}^{\prime\prime}$. The time it takes light to traverse
the sample is then
$t_{0}=\frac{L}{c}\sqrt{\epsilon _{0}^{^{\prime }}}$ (for a 1mm slab of Cu$_{2}$O with $%
\epsilon _{0}^{^{\prime }}=6.5$, $t_{0}=8.5$ps). If the time $t$
between the moment the light enters the sample and the moment it
is observed on the other side of the sample greatly exceeds
$t_{0}$, then one way to satisfy Eq.(\ref{omegan2}) would be to
consider processes, in which the light is reflected many times
($n\gg 1$) and passes a long way $\left( 2n+1\right) L$ before it
leaves the sample. Due to the rather strong absorption of visible
light in Cu$_{2}$O and the loss of intensity at each reflection
(the reflection coefficient $R^{2}\approx 0.2$), the intensity of
the transmitted light for $n\gg 1$ is extremely weak.

\begin{figure}[htbp]
\centering\includegraphics[width=\sfigwidth]{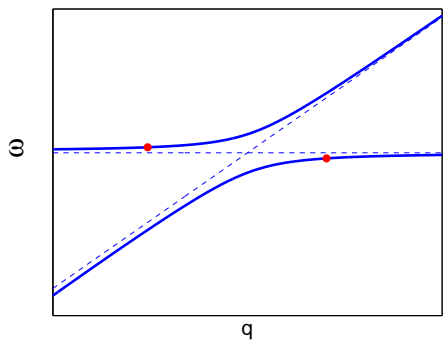}
\caption{\label{fig:polariton} Two branches of the
polariton dispersion (solid lines). The circles mark a
pair of polaritons moving with the same velocity.}
\end{figure}

Equation (\ref{omegan2}) can also be satisfied for small $n$, but
only if light in the sample can propagate slowly. This happens
when it is mixed with an exciton, i.e. when $\bar{\omega}\approx
\Omega $, where $\Omega $ is the exciton frequency. Such a mixing
results in two polariton branches $\omega _{\pm }\left( q\right)
$, both containing slowly propagating excitations (see
Fig.~\ref{fig:polariton}), so that Eq.(\ref{omegan2}) has two
solutions. Then Eq.(\ref{Eout4}) has to be modified and for each
$n$ the transmitted field is the sum of the contributions of the
two polariton trajectories:
\begin{equation}
E_{t} \approx 2E_{0}\sum_{n=0} \frac{\Delta _{n}}{\Delta
}\left( 1-R^{2}\right) R^{2n}e^{i\frac{\left(
S_{n+}+S_{n-}\right) }{2}}\cos \left(
\frac{S_{n+}-S_{n-}}{2}+\frac{\pi}{4}\right) ,
\label{Eout5}
\end{equation}
where $\pm$ labels the polariton branch. Below we show that 
for the quadrupolar of excitons to light
$\frac{S_{n+}-S_{n-}}{2}=\Omega \sqrt{\left( 2n+1\right)
ftt_{0}}$, where $f$ is the strength of the quadrupolar coupling,
so that the interference between the two polaritons moving with
the same velocity gives rise to oscillations of the intensity of
the transmitted light with a time-dependent period. Equation
(\ref{Eout5}) is actually an asymptotic expression for $E_{t}$
valid in the large-$t$ limit, when $\omega _{n+}-\omega _{n-}\gg
\Delta _{n}$, i.e. when the intervals of relevant frequencies
around the two stationary points do not overlap. A more general
expression is obtained below [see Eq.(\ref{Eout6})]. A similar
situation is found in scattering problems, when the interference
between different classical trajectories makes the scattering
cross-section a rapidly oscillating function of angle. A simple
semiclassical expression for the cross-section is not valid,
however, for angles close to $0$ and $\pi$, where the so-called
glory phenomenon takes place.\cite{glory}

The polariton dispersion is found from
\begin{equation}
\left( \omega ^{2}+i\omega \Gamma -\varepsilon
_{q}^{2}\right) \left( \omega ^{2}-\omega _{q}^{2}\right)
-f\omega ^{2}\omega _{q}^{2}=0,  \label{disp1}
\end{equation}
where $\omega _{q}=\frac{cq}{\sqrt{\epsilon _{0}}}$ and
$\varepsilon _{q}=\Omega +\frac{q2}{2M}$ are,
respectively, the photon and exciton frequencies, $f$ is
the strength of the quadrupolar photon-exciton coupling
and $\Gamma$ is the exciton decay width. Equation
(\ref{disp1}) corresponds to the dielectric
function\cite{Froelich,fro91}
\begin{equation}
\epsilon \left( \omega ,q\right) = \epsilon
_{0}+\frac{fc^{2}q^{2}}{ \varepsilon _{q}^{2}-\omega
^{2}-i\omega \Gamma }. \label{epsilon}
\end{equation}

It is convenient to introduce the dimensionless variables
$X=\frac{c^{2}q^{2}}{\Omega ^{2}\epsilon _{0}^{\prime }}$ (wave
vector squared) and $Y=\frac{\omega ^{2}}{\Omega ^{2}}$ (frequency
squared), which near the crossing point of the photon and exciton
dispersions are both close to 1. We also make use of the fact that
for Cu$_{2}$O the dimensionless parameters, describing the
exciton-photon coupling ($f\approx 4\cdot 10^{-9}$), exciton
kinetic energy ($\xi =\epsilon _{0}^{\prime} \frac{\Omega }{Mc^{2}
}\approx 10^{-5}$), decay width ($\gamma =\frac{\Gamma }{\Omega
}\approx 5\cdot 10^{-7}$),\cite{Froelich,fro91} and the absorption
of light $\left( \zeta =\frac{\epsilon
_{0}^{\prime\prime}}{\epsilon _{0}^{\prime}}\approx
10^{-4}\right)$ are rather small. Then Eq.(\ref{disp1}) can be
written in the form
\begin{equation}
\left( Y-1-\xi +i\gamma \right) \left( Y-X+i\zeta \right)
=f.  \label{disp2}
\end{equation}
Near the photon-exciton crossing point
\[
\left\{
\begin{array}{ccc}
X & = & 1+\xi + X^{\prime}, \\  Y & = & 1+\xi +Y^{\prime},
\end{array}
\right.
\]
with $X^{\prime},Y^{\prime}\ll 1$. Using Eq.(\ref{disp2}) we can
express the polariton `momentum' $X^{\prime}$ in terms of its
`frequency' $Y^{\prime}$:
\begin{equation}
X^{\prime}=Y^{\prime}-\frac{f}{Y^{\prime}+i\gamma }+i\zeta
. \label{X(Y)}
\end{equation}
Note that while $X^{\prime}\left( Y^{\prime}\right) $ is a
single-valued function, $Y^{\prime}\left(
X^{\prime}\right) $ has two branches.

\begin{figure}[htbp]
\centering\includegraphics[width=\sfigwidth]{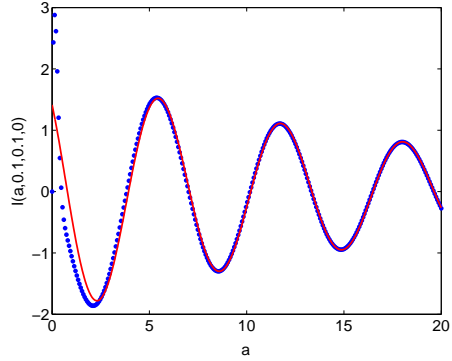}
\caption{\label{fig:integral} The $a$-dependence of the integral
$I(a,0.1,0.1,0)$ [see Eq.(\ref{I1})] calculated numerically (dots)
and in the stationary phase approximation (solid line).}
\end{figure}

The polariton action is then given by
\begin{equation}
S_{n}=-\omega t+q\left( 2n+1\right) L\approx \Omega \left[
\left( 2n+1\right) t_{0}-t\right] +\frac{\Omega
t}{2}\left[ r_{n}X^{\prime}-Y^{\prime}-\xi \right] ,
\label{Sn1}
\end{equation}
where $r_{n}=\left( 2n+1\right) \frac{t_{0}}{t}$ and, by
assumption, $r_{n}\ll 1$. Using the substitution
$Y^{\prime}=z\sqrt{fr_{n}}$ and Eq.(\ref{X(Y)}), we obtain
\begin{equation}
S_{n}\approx S_{n}^{\left( 0\right) }- \frac{\phi
_{n}}{2}\left[ z+ \frac{1}{z+i\frac{\gamma
}{\sqrt{fr_{n}}}}\right] . \label{Sn2}
\end{equation}
Here $S_{n}^{\left( 0\right) }=\Omega \left[ \left(
2n+1\right) t_{0}-t\right] +i\Omega \left( 2n+1\right)
t_{0}\frac{\zeta }{2}$ is the action for $f=0$ and
$\phi_{n}=\Omega t\sqrt{ fr_{n}}=\Omega \sqrt{\left(
2n+1\right) ftt_{0}}$ is the phase of the polariton
oscillations. Equation (\ref{Eout2}) can now be written in
the form
\begin{equation}
E_{t}\approx E_{0}\left( 1-R^{2}\right)
\sum_{n=0}\frac{\Delta _{n}}{\Delta
}R^{2n}e^{iS_{n}^{\left( 0\right)
}}I\left(\phi_n,\frac{\gamma
}{\sqrt{fr_{n}}},\frac{\phi_{n}}{2\Delta t},\frac{\xi
}{\sqrt{fr_{n}}}\right), \label{Eout6}
\end{equation}
where $\Delta _{n}^{2}=\frac{\Omega \sqrt{fr_{n}}}{4t}$,
$R=\frac{\sqrt{\epsilon _{0}^{\prime }}-1}{\sqrt{\epsilon
_{0}^{\prime }}+1}$, and
\begin{equation}
I(a,b,c,d)=\sqrt{\frac{a}{2\pi }}\int\limits_{-\infty
}^{+\infty }dze^{-i\frac{a}{2}\left( z+\frac{1}{z+ib}
\right) -\frac{c}{2}\left( z+d\right)^{2}} \label{I1}
\end{equation}
The factor $e^{-\frac{c}{2}\left( z+d\right) ^{2}}$ (with
$c=\left( \frac{\alpha _{n}}{2\Delta t}\right) ^{2}$ and
$d=\frac{\xi }{\sqrt{fr_{n}}}$), describing the shape of
the Gaussian wave packet with $\bar{\omega}=\Omega $, is
only important for $a \lesssim 1$. For $a\gg 1$,
$I(a,b,c,d)$ becomes independent of $c$ and $d$. In this
limit the integration over $z$ can be done in the
stationary-phase approximation, as the two stationary
points $z_{n\pm }$: $z_{n\pm }+i\beta _{n}=\pm 1$, are
sufficiently separated from each other. The result is
\begin{equation}
I(a,b,c,d)\approx 2e^{-\frac{ab}{2}}\cos \left(a+\frac{\pi
}{4}\right) . \label{I2}
\end{equation}
In Fig.~\ref{fig:integral} we compare $I(a,b,c,d)$, defined by
Eq.(\ref{I1}), for $b=c=0.1$ and $d=0$ (dots), with the asymptotic
expression Eq.(\ref{I2}) (solid line). We see that after about one
oscillation the two curves coincide and the transmitted field
equals the sum of the two polariton contributions
\begin{equation}
E_{t}\approx 2E_{0}\left( 1-R^{2}\right)
\sum_{n=0}R^{2n}e^{i\Omega \left[ \left( 2n+1\right)
t_{0}- t\right] -\frac{1}{2}\alpha
_{\Omega}(2n+1)L-\frac{\Gamma t}{2}}\frac{\Delta
_{n}}{\Delta }\cos \left( \Omega \sqrt{\left( 2n+1\right)
ftt_{0}}+\frac{\pi }{4}\right) \label{Eout7}
\end{equation}
[cf. Eq.(\ref{Eout5})], where $\alpha(\Omega)=\frac{\Omega
\epsilon _{0}^{\prime \prime }}{c\sqrt{\epsilon _{0}^{\prime }}}$
is the absorption coefficient at the frequency $\Omega$.

For further discussion it is useful to note that Eq.(\ref{Eout7})
can be written in the form
\begin{equation}
E_{t}\approx E_{0}\left( 1-R^{2}\right)
\sum_{n=0}R^{2n}F\left( l_{n},t\right) ,  \label{Eout8}
\end{equation}
where $F\left( l,t\right) $ is a function of the length
of the polariton path $l$ and the time of motion $t$:
\begin{equation}
F\left( l,t\right) = 2e^{i\Omega \left[
\frac{l\sqrt{\epsilon _{0}^{\prime }}}{c}-t\right]
-\frac{1}{2}\alpha(\Omega)l-\frac{\Gamma
t}{2}}\frac{\Delta \left( l\right) }{\Delta }\cos \left(
\Omega \sqrt{\frac{ftl\sqrt{\epsilon _{0}^{\prime
}}}{c}}+\frac{\pi }{4}\right) ,  \label{F}
\end{equation}
$l_{n}=$ $\left( 2n+1\right) L$ and $\Delta \left(
l\right) = \Delta _{0} \left[ \frac{l}{L}\right]
^{\frac{1}{4}}$.

\begin{figure}[htbp]
\centering\includegraphics[width=\sfigwidth]{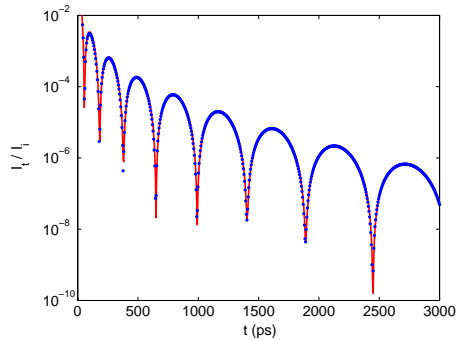}
\caption{\label{fig:It} The ratio of the intensities of the
transmitted and incident light for the parameters taken from
Ref.~[\onlinecite{Froelich,fro91}] calculated numerically using
Eq.(\ref{Eout6}) (dots) and in the stationary phase approximation
(solid line).}
\end{figure}

The ratio of the intensities of the transmitted and incident light
for the parameters taken from Ref.~[\onlinecite{Froelich,fro91}]
is shown in Fig.~\ref{fig:It}. The results of the numerical
calculation of Eqs.(\ref{Eout6}) and (\ref{I1})] are plotted with
dots, while the solid line was obtained using the semiclassical
expression Eq.(\ref{Eout7}). The difference between the two is
hardly visible in this semi-logarithmic plot, even for short times
$\sim 30$ ps. For the rather thick sample used in that experiment
($L = 0.91$mm) the contributions of the processes with multiple
internal reflections are negligibly small:
$\frac{I_t(n=1)}{I_t(n=0)} \approx 2 \cdot 10^{-3}$, so that the
transmitted intensity shows a single-mode oscillation, the period
of which grows with time.

\section{Polariton oscillations in non-linear process}
\label{sec:nonlinear}

In this section we return to the present experiments and apply the
semiclassical formalism discussed in the previous section to a
non-linear process, in which two photons with the frequency
$\omega_0$ are converted into a photon with the frequency close to
the ortho-exciton frequency $\Omega < 2\omega_0$. Our aim is to
explain the oscillations of the emitted light shown
Fig.\ref{oscillations}. Although small in amplitude and somewhat
irregular, these oscillations occur on the same time scale (of a
few hundred picoseconds) as the oscillations of the transmitted
light, discussed in Sec.\ref{sec:transmission}. Furthermore, the
inset in Fig.\ref{oscillations} shows the phase of these
oscillations grows proportionally to square root of time, just as
in the transmission experiment. Clearly, the intensity
oscillations observed in our experiment result from the polariton
beating.

This conclusion is actually very surprising, since the polariton
effects seem to be incompatible with the high-density exciton gas,
achieved in the present set-up. The exciton collision time in a
gas with the density $n=10^{19}$cm$^{-3}$ and temperature $T=30$K
is $\tau \sim 0.06$ps, which is three orders of magnitude smaller
than the oscillation period $\sim 500$ps. Furthermore, the
polariton momentum $p_{p}=\sqrt{\epsilon _{0}}\frac{\Omega }{c}$
is much smaller than the width of the thermal distribution in the
exciton gas $p_{T}=\sqrt{3MT}$: $p_{p}\sim 0.1p_{T}$. Thus the
elastic collisions effectively suppress the polariton oscillations
by removing very fast the excitons from the momentum range, in
which their interaction with light is important. The intensity
oscillations may only come from polaritons propagating through the
regions where the exciton gas density is sufficiently low. In our
experimental set-up the incoming light falls at a small angle
$\sim 10^{\circ }?$ with respect to the surface normal, while the
aperture angle was made rather large $\sim 20^{\circ }?$ to
increase the signal-to-noise ratio. This allows polaritons,
created at the rims of the light spot, to reach detector without
passing through the dense exciton gas regions.  The fact that the
exciton cloud grows with time does not alter this conclusion.
While the polariton momentum is much smaller than the thermal
momentum, the polariton velocity $v_{p}\sim
\frac{c}{\sqrt{\epsilon _{0}}}\frac{t_{0}}{t}$, where $t\sim 2$ns
is the time of the polariton motion, is much larger than the
thermal velocity $v_{T}=\sqrt{\frac{3T}{M}}$: $v_{p}\sim 20v_{T}$.
Furthermore, the length of the polariton path $\sim L$ covered
during that time is  much larger than increase of the size of the
exciton cloud  $\sim \sqrt{6v_{T}^{2}t\tau}$, so that polaritons
can  escape the expanding exciton cloud.

\begin{figure}[htbp]
\centering
\centering\includegraphics[width=\sfigwidth]{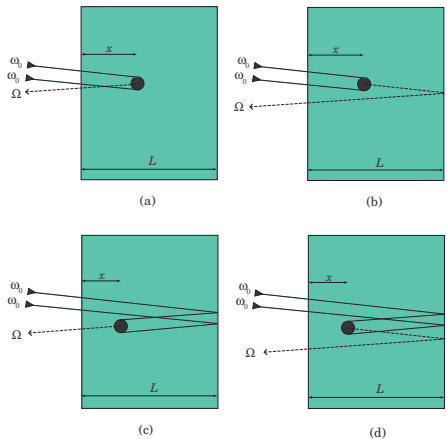}
\caption{\label{fig:process} Examples of nonlinear
processes, in which two photons with frequency $\omega_0$
are converted into a polariton with frequency $\Omega$.
For clarity, the directions of propagation of the incident
and emitted light are drawn slightly non-parallel.}
\end{figure}

The essential difference between the non-linear and transmission
experiments is the energy and momentum dissipation, which occurs
when the two in-coming photons are converted into a slowly
propagating polariton. As details of this process are poorly
understood, we simply assume here that the conversion is fast (on
the scale of the polariton oscillations) and that it occurs
locally, i.e. the polaritons are almost immediately emitted from
the spot, where the two photons were absorbed. Furthermore, we
assume that after the short thermalization time the motion of slow
polaritons (with the frequency close to $\Omega$) is essentially
free, i.e. their interactions with phonons and other excitons can
be neglected. In the transmission experiment\cite{Froelich,fro91}
the frequency of the incident light was deliberately chosen to be
close to the exciton frequency $\Omega$ to maximize the amplitude
of the polariton oscillations. In the nonlinear experiment one
obtains an exciton gas with a broad thermal distribution. The
energy and momentum selection of the excitons occurs in this case
automatically, since only the excitons with the energy close to
$\hbar \Omega$ are strongly mixed with light and contribute to the
polariton oscillations, and only those moving in the directions
approximately parallel and antiparallel to the sample surface
normal have a chance to be detected (examples of such processes
are shown in Fig.\ref{fig:process}). This selection, however,
strongly reduces the relative magnitude of the polariton
oscillations. A further reduction comes from the fact that the
energy and momentum dissipation destroys the interference between
the polaritons created at different points in the sample. In other
words, a luminescent exciton interferes only `with itself'. It is
surprising that even in this case the quantum polariton
oscillations are not entirely suppressed and can be detected.

In our simple model the intensity of the emitted light can
be calculated as follows. Using the semiclassical method
discussed above, we first find the radiation from the
polaritons created at a point with the coordinate $x$ in
the sample. Then this intensity is multiplied by the
conversion probability, weighted with a factor describing
the absorption of the incident light in the sample, and
integrated over $x$ from $0$ to $L$.

Consider, e.g., the process shown in Fig.~\ref{fig:process}a, in
which two incident photons with frequency $\omega _{0}$ enter the
sample (with the probability $\left( 1-R\right) ^{4}$), the
polariton travels along the path of length $x$ in the direction
opposite to that of the incident light, which gives the factor
$F^2\left( x,t\right)$ [cf. Eq.(\ref{Eout8})], and then leaves the
sample (the corresponding probability is $\left( 1+R\right)^{2}$).
For simplicity we assume the dielectric function for $f=0$ to be
frequency independent, so that the reflection of both incident and
emitted photons is described by the same coefficient $R$. The
intensity of this process is then given by
\begin{equation}
I_{a}\propto \left( 1-R\right) ^{4}\left( 1+R\right)
^{2}\int\limits_{0}^{L}dxe^{-2\alpha(\omega_0)x}F^2\left(x,t\right)
. \label{Ia}
\end{equation}
Here the factor $e^{-2\alpha(\omega_0)x}$ describes the
suppression of the intensity of the incident light
squared($\alpha(\omega_0) \approx
\frac{\omega_0}{\Omega}\alpha(\Omega)$). Actually, the
time of the polariton motion is somewhat shorter than the
total time $t$ between the moments light enters and leaves
the sample. However, the conversion is, by assumption,
fast and the time of motion of the incident photons in the
sample of thickness $L=0.3$mm, used in this experiment, is
only a few picoseconds, whereas the oscillation period is
of the order of one nanosecond.

In the process shown in Fig.~\ref{fig:process}b, the
polariton continues to move in the direction of the
incident light and is reflected from the right boundary
before leaving the sample (the path length is $2L-x$). The
reflection results in the extra factor $R^2$:
\begin{equation}
I_{b}\propto \left( 1-R\right) ^{4}R^{2}\left( 1+R\right)
^{2}\int\limits_{0}^{L}dxe^{-2\alpha(\omega_0)x}F^{2}\left(
2L-x,t\right). \label{Ib}
\end{equation}
Similarly, the sum of the intensities of the processes
shown in Fig.~\ref{fig:process}c and d is
\begin{equation}
I_{c} + I_{d} \propto \left( 1-R\right) ^{4}R^{4}\left(
1+R\right)
^{2}\int\limits_{0}^{L}dxe^{-2\alpha(\omega_0)(2L-x)}\left[F^{2}\left(
x,t\right) + R^2 F^{2}\left( 2L-x,t\right)\right].
\label{Icd}
\end{equation}
In the same way one can find the emission intensity for
the processes with multiple internal reflections of the
incident and emitted photons.

This simple approach gives a reasonable description of the
observed intensity oscillations, as shown in
Fig.~\ref{oscillations}, where we plot the oscillating part of
the intensity, $I_{osc} = I e^{\Gamma t}$, measured experimentally
(circles) and calculated theoretically. The dashed line is the sum
of the contributions of all four types of processes depicted in
Fig.~\ref{fig:process} with multiple reflections included. We
noted that the agreement with experiments improves, if the
contributions of the processes shown in Fig.~\ref{fig:process} a
and d, in which the polariton is emitted in the direction opposite
to that of incident light, are suppressed by the factor of 3. The
rationale for doing this may be the fact that not all polaritons
can reach the detector, since some of them are absorbed in the
regions with dense exciton gas. Therefore, each of the processes
discussed above has to be multiplied by a geometrical factor equal
the fraction of excitons that can be detected. We did not perform
the extensive calculations of this factor, since the results shown
in Fig.~\ref{oscillations} clearly show that the intensity
oscillations result from the polariton beatings.

\section{Conclusion}
In conclusion, the here presented B-TPE process provides an
efficient method to create a high density, cold exciton gas in the
bulk of cuprous oxide. The low initial temperature of the gas (despite the
energy mismatch), the instantaneous observation of \xo\
luminescence, and the preferential occupation of the $k\approx 0$
state seem to point to a coherent type of B-TPE process as
opposed to simple two photon absorption, and subsequent incoherent
decay to the ortho exciton state.
One possible process which would at least explain the low
initial gas temperature would be two photon absorption
from the upper valence band ($^{+}\Gamma_{7}$) that excites
a continuum of excitonic $p$ states associated with the
upper conduction band ($^{-}\Gamma_{8}$).\cite{rus73} Due
to the spin selection rules the excitons created this way
will be triplet excitons. However the $C_2$ band
originates from p orbitals so we do not have pure spin
states for the electron anymore. Therefore the angular
momentum selection rules allow the $C_2$ triplet exciton
to decay radiatively (dipole allowed transition) to either
triplet or singlet $C_1$ exciton and this way both para
and ortho states of a yellow series may be populated. This
could actually even happen in a coherent process
(hyper-Raman scattering), with the V1-C2 electron-hole
pair as the intermediate state.
The details of the B-TPE process remain unclear at the present. Future
experiments on photon emission coming from the C2-C1 transition
should yield the necessary information.

We gave simple analytical expressions for the polariton
oscillations observed both in the transmitted light in case of the
resonant excitation and in the emitted light in time resolved two
photon experiment. While in the first case the result is obtained
by a straightforward application of the semiclassical method, the
description of the beating that originates from a non-linear
process requires additional assumptions. Since we have not found
any evidence for the coherent production of the ortho-excitons, we
assumed that the coherence of the incident light is lost in the
irreversible conversion of two photons into a low-energy exciton.
We argue that this does not completely suppress the polariton
oscillations, since some polaritons can avoid the areas of dense
exciton gas on their way to detector and live long enough to be
considered as free particles. After such polaritons are created,
they propagate as quantum superpositions of two polaritons, which
gives rise to quantum beating, similar to the strangeness
oscillations in K-mesons produced in hadron reactions .[A.
Angelopoulos {\em et al.}, Phys. Lett. B {\bf 503}, 49 (2001).]
The shape of the oscillating intensity of light obtained within
this simple approach is in good agreement with the experimental
data.

Although the present study shows that ortho-exciton BEC
can not be achieved in this kind of experiments, mainly
because the particle decay rate is faster than the
temperature decay, the TPE method does provide an
efficient method to create a cold exciton gas. In
particular, it allows to create a para-exciton gas in
conditions where ortho-para down-conversion is blocked
({\em i.e.} at low temperatures), opening new
possibilities in the excitonic BEC quest.

{\em Acknowledgments. }
The authors acknowledge fruitful discussions with D. van der Marel.
This work was supported by the Stichting voor Fundamenteel
Onderzoek der Materie (FOM, financially supported by the
Nederlandse Organisatie voor Wetenschappelijk Onderzoek (NWO)).

\bibliography{cu2o}

\begin{thebibliography}{28}
\expandafter\ifx\csname natexlab\endcsname\relax\def\natexlab#1{#1}\fi
\expandafter\ifx\csname bibnamefont\endcsname\relax
  \def\bibnamefont#1{#1}\fi
\expandafter\ifx\csname bibfnamefont\endcsname\relax
  \def\bibfnamefont#1{#1}\fi
\expandafter\ifx\csname citenamefont\endcsname\relax
  \def\citenamefont#1{#1}\fi
\expandafter\ifx\csname url\endcsname\relax
  \def\url#1{\texttt{#1}}\fi
\expandafter\ifx\csname urlprefix\endcsname\relax\def\urlprefix{URL }\fi
\providecommand{\bibinfo}[2]{#2}
\providecommand{\eprint}[2][]{\url{#2}}

\bibitem[{\citenamefont{Moskalenko and Snoke}(2000)}]{mos00}
\bibinfo{author}{\bibfnamefont{S.~A.} \bibnamefont{Moskalenko}}
  \bibnamefont{and} \bibinfo{author}{\bibfnamefont{D.~W.} \bibnamefont{Snoke}},
  \emph{\bibinfo{title}{Bose-Einstein Condensation of Excitons}}
  (\bibinfo{publisher}{Cambridge University Press},
  \bibinfo{address}{Cambridge}, \bibinfo{year}{2000}).

\bibitem[{\citenamefont{Griffin et~al.}(1995)\citenamefont{Griffin, Snoke, and
  Stringari}}]{gri95}
\bibinfo{editor}{\bibfnamefont{A.}~\bibnamefont{Griffin}},
  \bibinfo{editor}{\bibfnamefont{D.~W.} \bibnamefont{Snoke}}, \bibnamefont{and}
  \bibinfo{editor}{\bibfnamefont{S.}~\bibnamefont{Stringari}}, eds.,
  \emph{\bibinfo{title}{Bose-Einstein Condensation}}
  (\bibinfo{publisher}{Cambridge University Press},
  \bibinfo{address}{Cambridge}, \bibinfo{year}{1995}).

\bibitem[{\citenamefont{Snoke and Negoita}(2000)}]{sno00}
\bibinfo{author}{\bibfnamefont{D.~W.} \bibnamefont{Snoke}} \bibnamefont{and}
  \bibinfo{author}{\bibfnamefont{V.}~\bibnamefont{Negoita}},
  \bibinfo{journal}{Phys. Rev. B} \textbf{\bibinfo{volume}{61}},
  \bibinfo{pages}{2904} (\bibinfo{year}{2000}).

\bibitem[{\citenamefont{Lai et~al.}(2004)\citenamefont{Lai, Zoch, Gossard, and
  Chemla}}]{lai04}
\bibinfo{author}{\bibfnamefont{C.~W.} \bibnamefont{Lai}},
  \bibinfo{author}{\bibfnamefont{J.}~\bibnamefont{Zoch}},
  \bibinfo{author}{\bibfnamefont{A.~C.} \bibnamefont{Gossard}},
  \bibnamefont{and} \bibinfo{author}{\bibfnamefont{D.~S.}
  \bibnamefont{Chemla}}, \bibinfo{journal}{Science}
  \textbf{\bibinfo{volume}{303}}, \bibinfo{pages}{503} (\bibinfo{year}{2004}).

\bibitem[{\citenamefont{Snoke et~al.}(2003)\citenamefont{Snoke, Liu, Denev,
  Pfeiffer, and West}}]{sno03}
\bibinfo{author}{\bibfnamefont{D.~W.} \bibnamefont{Snoke}},
  \bibinfo{author}{\bibfnamefont{Y.}~\bibnamefont{Liu}},
  \bibinfo{author}{\bibfnamefont{S.}~\bibnamefont{Denev}},
  \bibinfo{author}{\bibfnamefont{L.}~\bibnamefont{Pfeiffer}}, \bibnamefont{and}
  \bibinfo{author}{\bibfnamefont{K.}~\bibnamefont{West}},
  \bibinfo{journal}{Solid State Comm.} \textbf{\bibinfo{volume}{127}},
  \bibinfo{pages}{187} (\bibinfo{year}{2003}).

\bibitem[{\citenamefont{Snoke et~al.}(1991)\citenamefont{Snoke, Braun, and
  Cardona}}]{sno91}
\bibinfo{author}{\bibfnamefont{D.~W.} \bibnamefont{Snoke}},
  \bibinfo{author}{\bibfnamefont{D.}~\bibnamefont{Braun}}, \bibnamefont{and}
  \bibinfo{author}{\bibfnamefont{M.}~\bibnamefont{Cardona}},
  \bibinfo{journal}{Phys. Rev. B} \textbf{\bibinfo{volume}{44}},
  \bibinfo{pages}{2991} (\bibinfo{year}{1991}).

\bibitem[{\citenamefont{Naka and Nagasawa}(2002)}]{nak02}
\bibinfo{author}{\bibfnamefont{N.}~\bibnamefont{Naka}} \bibnamefont{and}
  \bibinfo{author}{\bibfnamefont{N.}~\bibnamefont{Nagasawa}},
  \bibinfo{journal}{Phys. Rev. B} \textbf{\bibinfo{volume}{65}},
  \bibinfo{pages}{075209} (\bibinfo{year}{2002}).

\bibitem[{\citenamefont{Naka and Nagasawa}(2004)}]{nak04}
\bibinfo{author}{\bibfnamefont{N.}~\bibnamefont{Naka}} \bibnamefont{and}
  \bibinfo{author}{\bibfnamefont{N.}~\bibnamefont{Nagasawa}},
  \bibinfo{journal}{Phys. Rev. B} \textbf{\bibinfo{volume}{70}},
  \bibinfo{pages}{155205} (\bibinfo{year}{2004}).

\bibitem[{\citenamefont{Kubouchi et~al.}(2005)\citenamefont{Kubouchi, Yoshioka,
  Shimano, Mysyrowicz, and Kuwata-Gonokami}}]{kub05}
\bibinfo{author}{\bibfnamefont{M.}~\bibnamefont{Kubouchi}},
  \bibinfo{author}{\bibfnamefont{K.}~\bibnamefont{Yoshioka}},
  \bibinfo{author}{\bibfnamefont{R.}~\bibnamefont{Shimano}},
  \bibinfo{author}{\bibfnamefont{A.}~\bibnamefont{Mysyrowicz}},
  \bibnamefont{and}
  \bibinfo{author}{\bibfnamefont{M.}~\bibnamefont{Kuwata-Gonokami}},
  \bibinfo{journal}{Phys. Rev. Lett.} \textbf{\bibinfo{volume}{94}},
  \bibinfo{pages}{016403} (\bibinfo{year}{2005}).

\bibitem[{\citenamefont{Fr\"{o}hlich et~al.}(1992)\citenamefont{Fr\"{o}hlich,
  Kulik, Uebbing, Langer, Stolz, and von~der Osten}}]{Froelich}
\bibinfo{author}{\bibfnamefont{D.}~\bibnamefont{Fr\"{o}hlich}},
  \bibinfo{author}{\bibfnamefont{A.}~\bibnamefont{Kulik}},
  \bibinfo{author}{\bibfnamefont{B.}~\bibnamefont{Uebbing}},
  \bibinfo{author}{\bibfnamefont{V.}~\bibnamefont{Langer}},
  \bibinfo{author}{\bibfnamefont{H.}~\bibnamefont{Stolz}}, \bibnamefont{and}
  \bibinfo{author}{\bibfnamefont{W.}~\bibnamefont{von~der Osten}},
  \bibinfo{journal}{Phys. Stat. Sol. (b)} \textbf{\bibinfo{volume}{173}},
  \bibinfo{pages}{31} (\bibinfo{year}{1992}).

\bibitem[{\citenamefont{{Fr{\"o}hlich}
  et~al.}(1991)\citenamefont{{Fr{\"o}hlich}, Kulik, and Uebbing}}]{fro91}
\bibinfo{author}{\bibfnamefont{D.}~\bibnamefont{{Fr{\"o}hlich}}},
  \bibinfo{author}{\bibfnamefont{A.}~\bibnamefont{Kulik}}, \bibnamefont{and}
  \bibinfo{author}{\bibfnamefont{B.}~\bibnamefont{Uebbing}},
  \bibinfo{journal}{Phys. Rev. Lett.} \textbf{\bibinfo{volume}{67}},
  \bibinfo{pages}{2343} (\bibinfo{year}{1991}).

\bibitem[{\citenamefont{Snoke et~al.}(1987)\citenamefont{Snoke, Wolfe, and
  Mysyrowicz}}]{sno87}
\bibinfo{author}{\bibfnamefont{D.~W.} \bibnamefont{Snoke}},
  \bibinfo{author}{\bibfnamefont{J.~P.} \bibnamefont{Wolfe}}, \bibnamefont{and}
  \bibinfo{author}{\bibfnamefont{A.}~\bibnamefont{Mysyrowicz}},
  \bibinfo{journal}{Phys. Rev. Lett.} \textbf{\bibinfo{volume}{59}},
  \bibinfo{pages}{827} (\bibinfo{year}{1987}).

\bibitem[{\citenamefont{Ruiz et~al.}(1997)\citenamefont{Ruiz, Alvarez, Alemany,
  and Evarestov}}]{rui97}
\bibinfo{author}{\bibfnamefont{E.}~\bibnamefont{Ruiz}},
  \bibinfo{author}{\bibfnamefont{S.}~\bibnamefont{Alvarez}},
  \bibinfo{author}{\bibfnamefont{P.}~\bibnamefont{Alemany}}, \bibnamefont{and}
  \bibinfo{author}{\bibfnamefont{R.~A.} \bibnamefont{Evarestov}},
  \bibinfo{journal}{Phys. Rev. B} \textbf{\bibinfo{volume}{56}},
  \bibinfo{pages}{7189} (\bibinfo{year}{1997}).

\bibitem[{\citenamefont{{O'H}ara and Wolfe}(2000)}]{oha00}
\bibinfo{author}{\bibfnamefont{K.~E.} \bibnamefont{{O'H}ara}} \bibnamefont{and}
  \bibinfo{author}{\bibfnamefont{J.~P.} \bibnamefont{Wolfe}},
  \bibinfo{journal}{Phys. Rev. B} \textbf{\bibinfo{volume}{62}},
  \bibinfo{pages}{12909} (\bibinfo{year}{2000}).

\bibitem[{\citenamefont{Denev and Snoke}(2002)}]{den02}
\bibinfo{author}{\bibfnamefont{S.}~\bibnamefont{Denev}} \bibnamefont{and}
  \bibinfo{author}{\bibfnamefont{D.~W.} \bibnamefont{Snoke}},
  \bibinfo{journal}{Phys. Rev. B} \textbf{\bibinfo{volume}{65}},
  \bibinfo{pages}{085211} (\bibinfo{year}{2002}).

\bibitem[{\citenamefont{Teh and Weichman}(1983)}]{teh83}
\bibinfo{author}{\bibfnamefont{C.~K.} \bibnamefont{Teh}} \bibnamefont{and}
  \bibinfo{author}{\bibfnamefont{F.~L.} \bibnamefont{Weichman}},
  \bibinfo{journal}{Can. J. Phys.} \textbf{\bibinfo{volume}{61}},
  \bibinfo{pages}{1423} (\bibinfo{year}{1983}).

\bibitem[{\citenamefont{Schmidt-Withley
  et~al.}(1974)\citenamefont{Schmidt-Withley, Martinez-Clemente, and
  Revcolevschi}}]{sch74}
\bibinfo{author}{\bibfnamefont{R.~D.} \bibnamefont{Schmidt-Withley}},
  \bibinfo{author}{\bibfnamefont{M.}~\bibnamefont{Martinez-Clemente}},
  \bibnamefont{and}
  \bibinfo{author}{\bibfnamefont{A.}~\bibnamefont{Revcolevschi}},
  \bibinfo{journal}{J. Crystal Growth} \textbf{\bibinfo{volume}{23}},
  \bibinfo{pages}{113} (\bibinfo{year}{1974}).

\bibitem[{\citenamefont{Ohyama et~al.}(1997)\citenamefont{Ohyama, Ogawa, and
  Nakata}}]{Ohyama}
\bibinfo{author}{\bibfnamefont{T.}~\bibnamefont{Ohyama}},
  \bibinfo{author}{\bibfnamefont{T.}~\bibnamefont{Ogawa}}, \bibnamefont{and}
  \bibinfo{author}{\bibfnamefont{H.}~\bibnamefont{Nakata}},
  \bibinfo{journal}{Phys. Rev. B} \textbf{\bibinfo{volume}{56}},
  \bibinfo{pages}{3871} (\bibinfo{year}{1997}).

\bibitem[{\citenamefont{Karpinska et~al.}(2005)\citenamefont{Karpinska, van
  Loosdrecht, Handayani, and Revcolevschi}}]{Karp}
\bibinfo{author}{\bibfnamefont{K.}~\bibnamefont{Karpinska}},
  \bibinfo{author}{\bibfnamefont{P.}~\bibnamefont{van Loosdrecht}},
  \bibinfo{author}{\bibfnamefont{I.}~\bibnamefont{Handayani}},
  \bibnamefont{and}
  \bibinfo{author}{\bibfnamefont{A.}~\bibnamefont{Revcolevschi}},
  \bibinfo{journal}{J. Lumin.} \textbf{\bibinfo{volume}{112}},
  \bibinfo{pages}{17} (\bibinfo{year}{2005}).

\bibitem[{\citenamefont{Gorban et~al.}(2000)\citenamefont{Gorban, Gubanov,
  Dmitruk, and Kulakovskii}}]{gor00}
\bibinfo{author}{\bibfnamefont{I.~S.} \bibnamefont{Gorban}},
  \bibinfo{author}{\bibfnamefont{V.~O.} \bibnamefont{Gubanov}},
  \bibinfo{author}{\bibfnamefont{I.~M.} \bibnamefont{Dmitruk}},
  \bibnamefont{and} \bibinfo{author}{\bibfnamefont{V.~D.}
  \bibnamefont{Kulakovskii}}, \bibinfo{journal}{J. Luminescence}
  \textbf{\bibinfo{volume}{87-89}}, \bibinfo{pages}{222}
  (\bibinfo{year}{2000}).

\bibitem[{\citenamefont{Ito and Masumi}(1997)}]{Ito}
\bibinfo{author}{\bibfnamefont{T.}~\bibnamefont{Ito}} \bibnamefont{and}
  \bibinfo{author}{\bibfnamefont{T.}~\bibnamefont{Masumi}},
  \bibinfo{journal}{J. Phys. Soc. Jpn.} \textbf{\bibinfo{volume}{66}},
  \bibinfo{pages}{2185} (\bibinfo{year}{1997}).

\bibitem[{\citenamefont{Rustagi et~al.}(1973)\citenamefont{Rustagi, Padere, and
  Mysyrowicz}}]{rus73}
\bibinfo{author}{\bibfnamefont{K.~C.} \bibnamefont{Rustagi}},
  \bibinfo{author}{\bibfnamefont{F.}~\bibnamefont{Padere}}, \bibnamefont{and}
  \bibinfo{author}{\bibfnamefont{A.}~\bibnamefont{Mysyrowicz}},
  \bibinfo{journal}{Phys. Rev. B} \textbf{\bibinfo{volume}{8}},
  \bibinfo{pages}{2721} (\bibinfo{year}{1973}).

\bibitem[{\citenamefont{Kavoulakis}(2002)}]{kav02}
\bibinfo{author}{\bibfnamefont{G.~M.} \bibnamefont{Kavoulakis}},
  \bibinfo{journal}{Phys. Rev. B} \textbf{\bibinfo{volume}{61}},
  \bibinfo{pages}{035204} (\bibinfo{year}{2002}).

\bibitem[{\citenamefont{Sun et~al.}(2002)\citenamefont{Sun, Rivkin, Chen,
  Ketterson, Markworth, and Chang}}]{sun02}
\bibinfo{author}{\bibfnamefont{Y.}~\bibnamefont{Sun}},
  \bibinfo{author}{\bibfnamefont{K.}~\bibnamefont{Rivkin}},
  \bibinfo{author}{\bibfnamefont{J.}~\bibnamefont{Chen}},
  \bibinfo{author}{\bibfnamefont{J.~B.} \bibnamefont{Ketterson}},
  \bibinfo{author}{\bibfnamefont{P.}~\bibnamefont{Markworth}},
  \bibnamefont{and} \bibinfo{author}{\bibfnamefont{R.~P.} \bibnamefont{Chang}},
  \bibinfo{journal}{Phys. Rev. B} \textbf{\bibinfo{volume}{66}},
  \bibinfo{pages}{245315} (\bibinfo{year}{2002}).

\bibitem[{\citenamefont{Lee and Zhu}(2000)}]{Lee}
\bibinfo{author}{\bibfnamefont{Y.~C.} \bibnamefont{Lee}} \bibnamefont{and}
  \bibinfo{author}{\bibfnamefont{W.}~\bibnamefont{Zhu}}, \bibinfo{journal}{J.
  Phys.: Condensed Matter} \textbf{\bibinfo{volume}{12}}, \bibinfo{pages}{L49}
  (\bibinfo{year}{2000}).

\bibitem[{\citenamefont{Ell et~al.}(1998)\citenamefont{Ell, Ivanov, and
  Haug}}]{ell}
\bibinfo{author}{\bibfnamefont{C.}~\bibnamefont{Ell}},
  \bibinfo{author}{\bibfnamefont{A.}~\bibnamefont{Ivanov}}, \bibnamefont{and}
  \bibinfo{author}{\bibfnamefont{H.}~\bibnamefont{Haug}},
  \bibinfo{journal}{Phys. Rev.B} \textbf{\bibinfo{volume}{57}},
  \bibinfo{pages}{9663} (\bibinfo{year}{1998}).

\bibitem[{\citenamefont{Haug}(1999)}]{haug}
\bibinfo{author}{\bibfnamefont{H.}~\bibnamefont{Haug}},
  \bibinfo{journal}{Journal of Lumin} \textbf{\bibinfo{volume}{69}},
  \bibinfo{pages}{83} (\bibinfo{year}{1999}).

\bibitem[{\citenamefont{Mott and Massey}(1965)}]{glory}
\bibinfo{author}{\bibfnamefont{N.}~\bibnamefont{Mott}} \bibnamefont{and}
  \bibinfo{author}{\bibfnamefont{H.}~\bibnamefont{Massey}},
  \emph{\bibinfo{title}{Theory of Atomic Collisions}}
  (\bibinfo{publisher}{Clarendon}, \bibinfo{address}{Oxford},
  \bibinfo{year}{1965}).

\end{thebibliography}
\end{document}